\documentclass[aps,preprint]{revtex4}%
\usepackage{amsfonts}
\usepackage{amsmath}
\usepackage{amssymb}
\usepackage{graphicx}%
\usepackage{xcolor}
\usepackage{braket}
\providecommand{\U}[1]{\protect\rule{.1in}{.1in}}

\begin{document}
\title[Short title for running header]{Unsupervised Feature Extraction and Reconstruction Using Parameterized Quantum Circuits}

\author{Li-An Lo}
\affiliation{Department of Physics, Chung Yuan Christian University, Chungli 32081, Taiwan}

\author{Li-Yi Hsu}
\affiliation{Department of Physics, Chung Yuan Christian University, Chungli 32081, Taiwan}
\affiliation{
Physics Division, National Center for Theoretical Sciences, Taipei 106319, Taiwan }

\author{En-Jui Kuo}
\affiliation{
Department of Electrophysics, National Yang Ming Chiao Tung University, Hsinchu, Taiwan, R.O.C.}
\pacs{PACS number}

\begin{abstract}
Autoencoders are fundamental tools in classical computing for unsupervised feature extraction, dimensionality reduction, and generative learning. The Quantum Autoencoder (QAE), introduced by Romero J. \cite{qae}, extends this concept to quantum systems and has been primarily applied to tasks like anomaly detection. Despite its potential, QAE has not been extensively evaluated in basic classification tasks such as handwritten digit classification, which could provide deeper insights into its capabilities.

In this work, we systematically investigate the performance of QAE in unsupervised feature extraction and reconstruction tasks. Using various encoder and decoder architectures, we explore QAE's ability to classify MNIST and analyze its reconstruction effectiveness. Notably, with a QCNN-based encoder, QAE achieves an average accuracy of 97.59$\%$ in binary classification of MNIST (0 and 1), demonstrating the feasibility of QAE for such tasks. This result highlights the synergy between QAE and QCNN in achieving optimal feature extraction performance. However, we also identify limitations in QAE’s reconstruction capability, underscoring the need for further advancements in encoder and decoder designs. Our findings provide a foundation for future research on leveraging QAE in practical quantum machine learning applications.
\end{abstract}
\volumeyear{year}
\volumenumber{number}
\issuenumber{number}
\eid{identifier}
\date[Date text]{date}
\received[Received text]{date}

\revised[Revised text]{date}

\accepted[Accepted text]{date}

\published[Published text]{date}

\maketitle
\tableofcontents

\section{INTRODUCTION}

Unsupervised feature extraction refers to an effective method in machine
learning that identifies and selects important features from unlabeled
datasets, and the original high-dimensional data are transformed into
informative lower-dimensional space, which is
crucial for processes such as image processing \cite{u1}, natural language
processing \cite{u2}, and bioinformatics \cite{u3}. To achieve this unsupervised learning tasks, the autoencoders as a
type of neural network are usually exploited. Schematically, an autoencoder primarily includes both an encoder and a
decoder. The encoder compresses the input into a lower-dimensional
representation while capturing its essential features. On the other hand, the
decoder is to reconstruct an approximation of the input from the compressed
representation. To minimize the difference between the input and the reconstructed output, the classical encoder and decoder are both fully connected
forward neural networks, enabling an autoencoder to reduce data's
dimensionality and extract the essential features for accurate reconstruction.
 Autoencoders have various applications, including dimensionality reduction \cite{dr}, feature learning \cite{fl}, data denoising \cite{dt}, anomaly detection \cite{ad}, generative tasks \cite{gt}, image classification \cite{ic}, and applications in condensed matter physics \cite{cft}. 

Convolutional autoencoders represent the amalgamation of convolution neural networks and traditional autoencoders \cite{ca1}. The convolutional neural networks (CNN) are forward neural networks used mostly for supervised learning, and well-known for image classification \cite{cnnic}. There are two fundamental
building blocks in convolutional neural networks. One is the convolution layers that extract the feature using convolutional operations. The other is the
pooling layers that reduce the dimensions of the feature maps and avoid overfitting. As for convolutional autoencoders, their architecture includes an encoder and a decoder, similar to that of autoencoders. The encoder achieves the compression and feature extraction of the input data by applying the convolutional and pooling layers, and the decoder employs the deconvolutions to reconstruct the original input with high accuracy. Convolutional autoencoders are exploited for the task of image denoising \cite{ca2}, feature extraction and transfer learning \cite{ca1}, medical image \cite{ca3,ca33}, anomaly detection \cite{ca4}, 3D object reconstruction \cite{ca5}.

Regarding the increasing size of datasets and the exponentially growing parameters
in large-scale tasks, quantum machine learning has recently attracted significant
research attention due to its potential computational advantage in solving
problems intractable on classical computers. In quantum computing,
quantum-mechanical phenomena such as superposition and entanglement in a large
Hilbert space allow for speed-ups over the limitations of classical
computation \cite{1,2,3}. Although fault-tolerant quantum computing is far
from physical realization in the foreseeable future, recent advancements in
quantum technology have enabled research efforts devoted to quantum machine
learning on noisy intermediate-scale quantum (NISQ) devices \cite{n1,n2,n3}. In quantum
machine learning, variational quantum algorithms (VQAs) within the hybrid
quantum-classical framework are extensively employed. These algorithms utilize
classical optimizers to train parameterized quantum circuits (PQCs) during the
learning process. However, designing the layout of parameterized quantum
circuits for a specific training task remains challenging. Considering the
unavoidable noise in physical implementations, the depth of quantum circuits
should be kept shallow, and the use of entangling two-qubit gates should be
minimized.

In this work, we employ variational quantum algorithms for unsupervised learning
tasks. We introduce quantum convolutional neural networks (QCNNs) for
unsupervised feature extraction. All convolutional and pooling
layers in the proposed QCNN comprise specific two-qubit unitary operations as
ansatz. Next, we study quantum convolutional autoencoders (QCAEs) for unsupervised
data classification. In this approach, the proposed QCNN and its unitary inverse
are exploited as the encoder and decoder, respectively. By comparison, the
quanvolutional neural networks in \cite{qcnn3, qca} comprise both
quantum and classical convolutional layers. Similarly, in the quantum convolutional
autoencoder proposed in \cite{qca1}, the encoder and decoder are both classical,
and a quantum circuit in between executes the quantum approximate
optimization algorithm (QAOA) \cite{QAOA}.

In Section II, we review related works on data embedding and parameterized quantum circuits (PQCs). Section III introduces the principles of Quantum Autoencoders (QAE). Section IV provides a detailed illustration of the proposed QCNN-based architecture and alternative PQCs for variational quantum algorithms (VQAs) in unsupervised feature extraction and data classification tasks. Section V elaborates on the quantum data encoding techniques employed in this study, while Section VI discusses the loss functions used. In Section VII, we present results on unsupervised feature extraction and reconstruction, alongside additional tests and the outcomes from real-device experiments. Finally, Section VIII concludes the paper.


\section{RELATED WORK}

The quantum version of autoencoders was initially introduced for efficient
compression of quantum data \cite{qae}. Later enhanced feature quantum
autoencoders were investigated for compressing ground states and classical
handwritten digits \cite{qae2}. Quantum autoencoders have their applicaitons
in quantum data denoising \cite{qae22,qae23,qae24,qae25}, error-mitigation
\cite{qae3}, error-correction \cite{qae5}, and leraning hard distribution
\cite{qae6,qae7}. In addition,\ various quantum convolution neural networks
attracted much more atterntion \cite{qcnn, bqcnn,qcnn2,qcnn1,qcnn8} and have
their applications in classical data classification \cite{aqcnn}, image
classification \cite{qcnn10,qcnn11,qcnn12,qcnn4}, data analysis in high energy
physics \cite{qcnn5}, damage and object detection \cite{qcnn6,qcnn9}.

\begin{figure}[h!]
    \includegraphics[width=16.4cm]{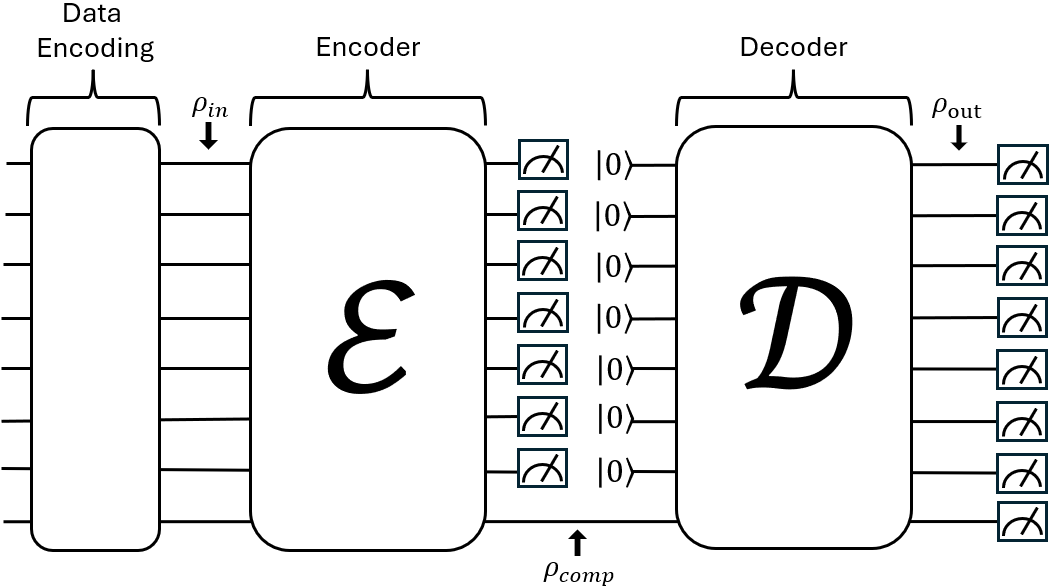}
    \caption{Graphical representation of the quantum convolutional
autoencoder. The unitary map $\mathcal{E}$ extracts features from the $n$-qubit
input state $\rho_{\text{in}}$ into the $k$-qubit compressed state $\rho_{\text{comp}}$.
The decoder $\mathcal{D} = \mathcal{E}^{\dagger}$ attempts to map $\rho_{\text{comp}}$
along with an additional $k$-qubit register initially prepared in the state 
$\left\vert 0\right\rangle ^{\otimes k}$ back to the $n$-qubit output state 
$\rho_{\text{out}}$, reconstructing $\rho_{\text{in}}$ All quantum channels are CPTP maps (completely positive (CP) trace-preserving).}
    \label{fig:StructureOfQAE}
\end{figure}

\begin{figure}[h!]
    \includegraphics[width=16.4cm]{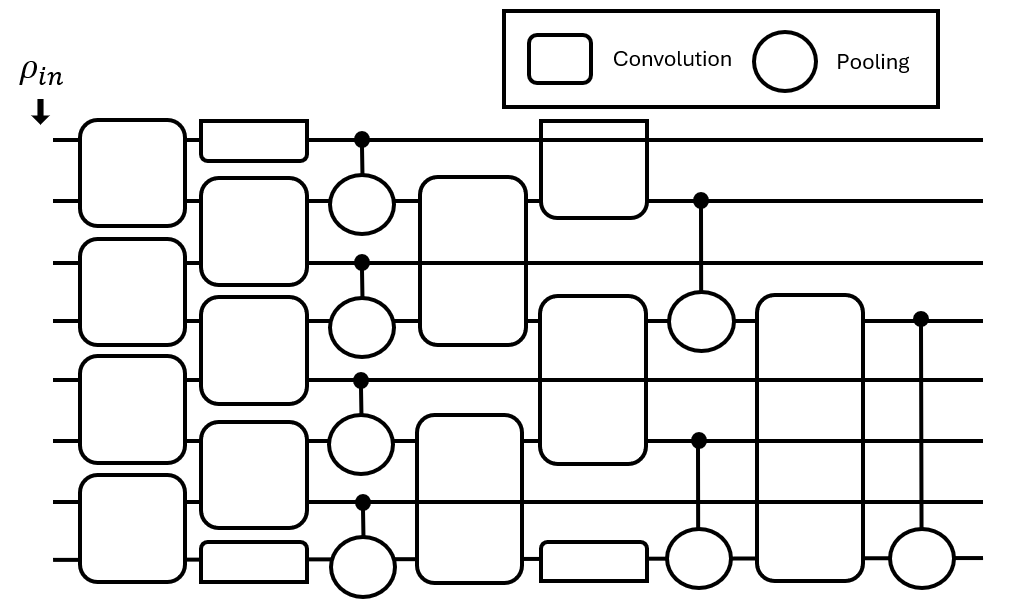}
    \caption{Circuit implementation of the quantum convolutional autoencoder with QCNN as encoder. We also call this Circuit as "C-P-C-P-C-P", meaning it uses three layers of Convolution and Pooling.}
    \label{fig:StructureOfQCNN}
\end{figure}

The schematic representation of the proposed quantum convolutional autoencoder and its
circuit implementation is pictorially depicted in Figure \ref{fig:StructureOfQAE}. The encoder is a
parametrized quantum circuit consisting of two-qubit ansatzes as convolutional
and pooling layers, as shown in Figure \ref{fig:StructureOfQCNN}. Notably, we pick one of the two-qubit
ansatz in Fig. 3 as the convolutional layer \cite{aqcnn}.

\section {Quantum Autoencoder}

In the entire quantum circuit reconstruction of the Quantum Autoencoder (QAE), given $n = c + t$ qubits as input, the encoder removes $t$ qubits, referred to as "trash qubits." The remaining $c$ qubits are termed "compressed qubits." After this process, $a$ "reference qubits" ($a=t$) are added and passed together with the compressed qubits, are passed through the decoder to reconstruct the output. Therefore, the complete QAE reconstruction process requires $c + a + t = c + 2t$ qubits. However, in unsupervised feature extraction, we only focus on training the encoder, which involves just $c + t$ qubits.

The working principle of the Quantum Autoencoder (QAE) is illustrated in Figure \ref{fig:StructureOfQAE}. The compressed and output density matrix can be expressed as:
\begin{align}
\rho_{\text{comp}} = \text{Tr}_t\big[\mathcal{E}(\rho_{\text{in}})\mathcal{E}^\dag\big],
\end{align}
\begin{align}
\rho_{\text{out}} = \mathcal{D}(\rho_{\text{comp}} \otimes \rho_{a})\mathcal{D}^\dag, \theta'),
\end{align}

Where $\text{Tr}_t(.)$ denotes the partial trace operation over the QAE’s trash qubits, and $\mathcal{E}$ represents a Completely Positive Trace-Preserving (CPTP) map, modeling the encoder transformation. The decoder (as a CPTP map D) performs the inverse transformation. 

Let $\{ \theta \}$ and $\{ \theta' \}$ be the parameter sets for the encoder and decoder, we can rewrite Equation (1) and (2) as:
\begin{align}
\rho_{\text{comp}} = \text{Tr}_t\big[\mathcal{E}(\rho_{\text{in}}, \theta)\big],
\end{align}
\begin{align}
\rho_{\text{out}} = \mathcal{D}(\rho_{\text{comp}}, \theta'),
\end{align}

For perfect reconstruction ($\rho_{\text{out}} = \rho_{\text{in}}$), the QAE must satisfy the following condition:
\begin{align} 
    \mathcal{D}\big(\text{Tr}_t\big[\mathcal{E}(\rho_{\text{in}}, \theta)\big]\otimes\rho_{a}, \theta'\big) = \rho_{\text{in}}.
\end{align}

Where $\rho_a = \ket{a}\bra{a}$ is the density matrix of the reference qubits, typically initialized as $\ket{a} = \ket{0}^{\otimes t}$.
Due to the reversibility of quantum circuits, the above condition can be equivalently rewritten as:
\begin{align} 
    \text{Tr}_t\big[\mathcal{E}(\rho_{\text{in}} ,\theta)\big]\otimes\rho_{a} = \mathcal{D}^{-1}(\rho_{\text{in}}, \theta') = \mathcal{E}(\rho_{\text{in}}, \theta').
\end{align}

Furthermore, we can express this relationship as:
\begin{align}
\mathcal{E}(\rho_{\text{in}}, \theta) \otimes \rho_{a} = \mathcal{E}(\rho_{\text{in}}, \theta') \otimes \rho_{t}.
\end{align}
Where $\rho_t = \ket{t}\bra{t}$ is the density matrix of the trash qubits.

Since this equation involves only the encoder, it follows that equality holds if both: $\theta = \theta'$ and $\ket{a} = \ket{t}$. In general, we cannot perfectly reconstruct the original state due to the loss of information in the encoding process. However, we can quantify the reconstruction error using a suitable distance metric. A common choice is the fidelity between the input and output states, given by  
\begin{align}  
\mathcal{F}(\rho_{\text{in}}, \rho_{\text{out}}) = \left( \text{Tr} \sqrt{\sqrt{\rho_{\text{in}}} \rho_{\text{out}} \sqrt{\rho_{\text{in}}}} \right)^2.
\end{align}  
Alternatively, we can measure the trace distance, defined as  
\begin{align}  
D_{\text{tr}}(\rho_{\text{in}}, \rho_{\text{out}}) = \frac{1}{2} \|\rho_{\text{in}} - \rho_{\text{out}}\|_1.
\end{align}  
These metrics allow us to evaluate the degree to which the reconstructed state retains information from the original state. Here $||\bullet||_1$ stands for the Schatten $1$ norm.

\section{Encoder Architecture}
\subsection{Ansatz-based Quantum Convolutional Neural Network (AQCNN)}
QCNNs, as a quantum analog of classical convolutional neural networks, were first introduced in \cite{qcnn}. The structure of QCNNs comprises convolutional and pooling layers, with each layer implemented as a two-qubit unitary operation, and their parameters and structure are the same.

The convolutional layers in the QCNN perform localized feature extraction by entangling nearby qubits, while the pooling layers reduce dimensionality by discarding part of the quantum state, allowing for a compressed representation of the original input. This approach enables efficient data processing on quantum hardware while capturing essential features.

\begin{figure}[h!]
    \includegraphics[width=16.4cm]{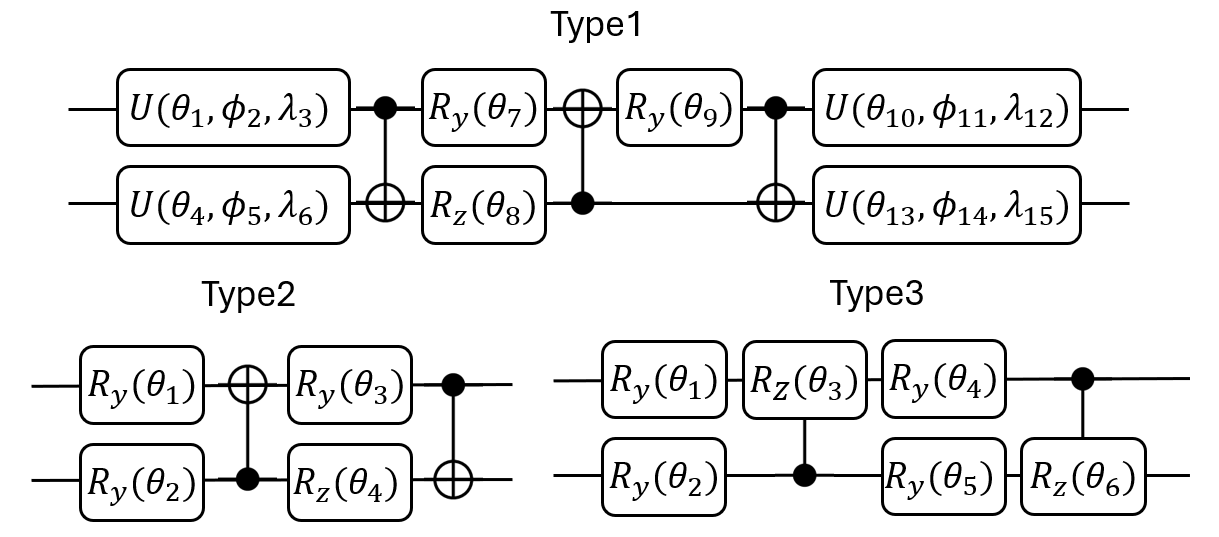}
    \caption{The convolutional structure we employed is illustrated here, showcasing three of the nine convolutional circuits proposed by T. Hur in \cite{aqcnn}. Among these, Type 1 achieves the best performance but requires a greater number of parameters. The unitary operator is defined as
$U(\theta, \phi, \lambda) = 
\begin{bmatrix}
\cos\left(\frac{\theta}{2}\right) & -e^{i\lambda}\sin\left(\frac{\theta}{2}\right) \\
e^{i\phi}\cos\left(\frac{\theta}{2}\right) & e^{i(\phi+\lambda)}\sin\left(\frac{\theta}{2}\right)
\end{bmatrix},$
and \( R_y(\theta) = \exp(-i \theta Y / 2) \) represents a rotation around the \( y \)-axis by an angle \(\theta\). Similarly, we define \( R_z(\theta) = \exp(-i\theta Z / 2) \) and \( R_x(\theta) = \exp(-i \theta X / 2) \), where \( X, Y, Z \) are the standard Pauli matrices.}
\label{fig:ConvolutionStructure}
\end{figure}

\begin{figure}[h!]
    \includegraphics[width=14.2cm]{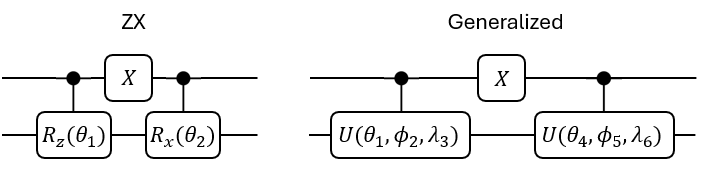}
    \caption{The pooling structure we used, utilizing two configurations. We call them "ZX" and "Generalized", as described in \cite{pool}.}

    \label{fig:PoolingStructure}
\end{figure}

In our proposed AQCNN, three parameterized convolutional ansatz and two pooling ansatz are used. The convolutional ansatzes (Type 1, 2, and 3, see Fig \ref{fig:ConvolutionStructure}) and the pooling layers (ZX pooling and generalized pooling, see Fig \ref{fig:PoolingStructure}) are optimized to balance feature extraction and computation cost.

\subsection{Other architecture}
\begin{figure}[h!]
    \includegraphics[width=14.2cm]{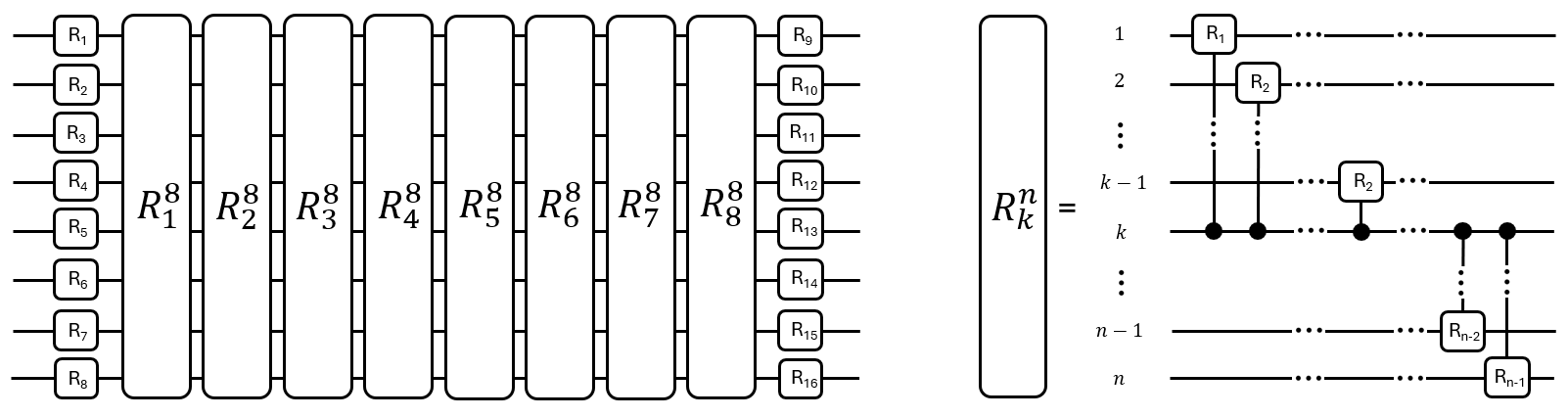}
    \includegraphics[width=14.2cm]{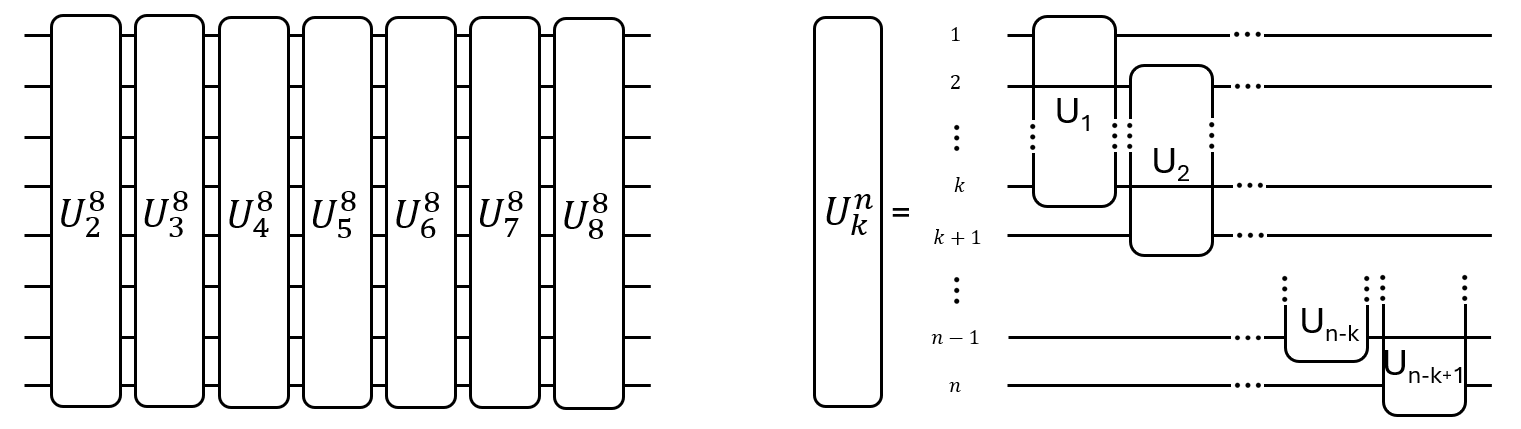}
    
    \caption{Two $n$-qubit programmable circuits used for unsupervised feature extraction and as the encoder of the autoencoder. (a) Architectures A (Above), with $n$ layers, the notation $R^n_k$ represents a series of controlled rotation gates (all three directions are all allowed), where all qubits except the k-th applies controlled single-qubit rotations on all other qubits.   (b) Architectures B (down), with $n-1$ layers, uses sequential two-qubit operations to entangle qubits for robust feature representation and capturing complex correlations in the data. The notation $U^n_k$ represents a series of any two-qubit operations applied to $n$ qubits, arranged from top to bottom with a spacing of $k-1$ qubits between each gate. It is evident that the parameter complexity of QCNN is $O(\log n)$, whereas Architectures A and B both have a complexity of $O(n^2)$.}
    \label{fig:StructureOfB}
\end{figure}

In addition to ansatz-based QCNNs, we use two other universal programmable quantum circuits for unsupervised learning tasks as comparison. We call them architecture A and architecture B introduced in \cite{pp1} and \cite{pp2}, respectively. As shown in Fig \ref{fig:StructureOfB}. Each of them comprises several layers and has a fixed network of gates with polynomially-scaled parameters. 

\section{Quantum Data Encoding}
In this work, two encoding methods, amplitude encoding and angle encoding, are employed to map classical data into quantum states.

\subsection{Amplitude Encoding}
Amplitude encoding represents classical data by mapping it into the amplitude of quantum states. In the dataset, each input has $N$ features, we use $n = \log_2(N)$ qubits. Given a classical data vector $\vec{x} = [x_1, x_2, \ldots, x_{N}]$, the amplitude encoding process maps $\vec{x}$ into a quantum state $\lvert \phi(\vec{x}) \rangle$ as follows:
\[
\lvert \phi(\vec{x}) \rangle = \sum_{i=1}^{N} \frac{x_i}{\|\vec{x}\|} \lvert i \rangle,
\]
where $\|\vec{x}\| = \sqrt{\sum_{i=1}^{N} |x_i|^2}$ is the Euclidean norm of $\vec{x}$, ensuring the state is normalized. This encoding allows us to store an exponentially large amount of data in a limited number of qubits, as each element $x_i$ is represented by the amplitude of the corresponding basis state $\lvert i \rangle$. However, state preparation can be complex, particularly for large datasets.

Amplitude encoding is highly efficient in terms of qubit usage, enabling large datasets to be encoded with relatively few qubits. This efficiency makes it appealing for quantum applications that require compact data representation. However, a significant disadvantage of amplitude encoding is that preparing an amplitude-encoded quantum state can be computationally demanding. The encoding process often requires complex gate sequences, which can be challenging to implement accurately on noisy intermediate-scale quantum (NISQ) devices.

\subsection{Angle Encoding}
Angle encoding represents data by associating classical values with rotation angles in quantum gates. For a data vector $\vec{x} = [x_1, x_2, \ldots, x_{N}]$, each feature $x_i$ is mapped to a rotation angle in a quantum circuit. Using $n$ qubits, the encoding process creates a quantum state $\lvert \phi \rangle$ with the following mapping:
\[  
\lvert \phi \rangle = \bigotimes_{i=1}^{N} R_y(x_i) \lvert 0 \rangle^N,
\]

where $R_y(x_i) = \exp(-i x_i Y / 2)$ is a rotation around the $y$-axis by angle $x_i$. We use similar notation for $R_z(x_i) = \exp(-i x_i Z / 2), R_x(x_i) = \exp(-i x_i X / 2)$ where $X,Y,Z$ are the standard pauli matrices.
This method allows for a straightforward implementation on quantum hardware, as each $x_i$ directly sets the rotation angle of a qubit. 

Angle encoding offers efficiency for noisy intermediate-scale quantum (NISQ) devices and is straightforward to implement, as it involves only single-qubit rotations determined by the data values. This simplicity in circuit design makes angle encoding well-suited to current quantum hardware. However, a notable drawback of angle encoding is that it typically requires a larger number of qubits compared to amplitude encoding for datasets of the same size. Each data point generally corresponds to an individual rotation, resulting in greater qubit demand as the dataset size increases.

Our simulation experimental results show that angle encoding performs better than amplitude encoding for unsupervised feature extraction, that is significantly different from the supervised binary classification study in \cite{qcnn}.


\section{Loss function}
In this work, we utilize two types of loss functions for training the quantum model: Mean Squared Error (MSE) Loss \cite{MS} and Binary Cross Entropy (BCE) Loss \cite{Ruby}. The MSE loss function measures the difference between the renferece qubits \( y \) and the model's predicted probability \( ||\braket{1|\mathcal{E}(\ket{\phi(\vec{x})},\theta)}||^2 \), defined as:
\begin{align} 
    MSELoss(\theta) 
    &= \frac{1}{T} \sum^{T}_{i=1}(y_i - ||\braket{1|\mathcal{E}(\ket{\phi(\vec{x})},\theta)_i}||^2)^2 
\end{align}
where $\mathcal{E}$ is the encoder with parameters $\theta$ which mapping $n$ qubits states into $k$ qubits state. $T$ is the number of the trash qubits states . 

This loss is particularly useful when dealing with continuous-valued targets, aiming to minimize the squared difference between the target and predicted values.

For binary classification tasks, we employ the Binary Cross Entropy (BCE) loss, which penalizes incorrect predictions more effectively for binary outputs. BCE loss is formulated as:
\begin{align} 
    BCELoss(\theta) 
    &=  -\frac{1}{T}\sum^{T}_{i=1} \left( y_i \log(||\braket{1|\mathcal{E}(\ket{\phi(\vec{x})},\theta)_i}||^2) + (1 - y_i) \log(||\braket{0|\mathcal{E}(\ket{\phi(\vec{x})},\theta)_i}||^2) \right)
\end{align}

This function penalizes predictions based on the logarithmic likelihood, ensuring that probabilities close to the true binary class values are rewarded, while incorrect predictions are penalized more heavily. Both loss functions help optimize the model parameters \(\theta\) for accurate feature extraction and classification within the quantum neural network.



\bigskip

\section{Results}
\subsection{Dataset}
We conduct experiments on the MNIST dataset, a widely-used collection of handwritten digits for testing feature extraction techniques \cite{mnist}. The dataset serves as an excellent benchmark for assessing the performance of quantum encoding methods and unsupervised learning models due to its high-dimensional structure and complex patterns.

For data preprocessing, if the input data uses Angle Encoding, the original data is first reduced in dimensionality using a classical autoencoder. On the other hand, if the input data uses Amplitude Encoding, the original image data is resized to dimensions of \(16 \times 16 = 256\) to match the required format.

\subsection{Training}
We randomly selected 400 samples from the test dataset to calculate accuracy. Training simulation using 8 qubits and compressed to 1 qubit with a batch size of 32 and a fixed learning rate of 0.005, and use the SGD(Stochastic gradient descent) optimization algorithm, which means that each optimization uses a random mini-batch of samples to optimize the parameters. Each training consisted of 64 iterations, and repeat different trainings to obtain accuracy results and calculate the average accuracy. For the main results, we consistently employed Angle Encoding, Type1 Convolution, ZX Pooling, and the BCE Loss Function to compare the unsupervised feature extraction and reconstruction capabilities across different architectures.

\subsection{Rotate The Bloch Sphere}
\begin{figure}[h!]
    \includegraphics[width=14.2cm]{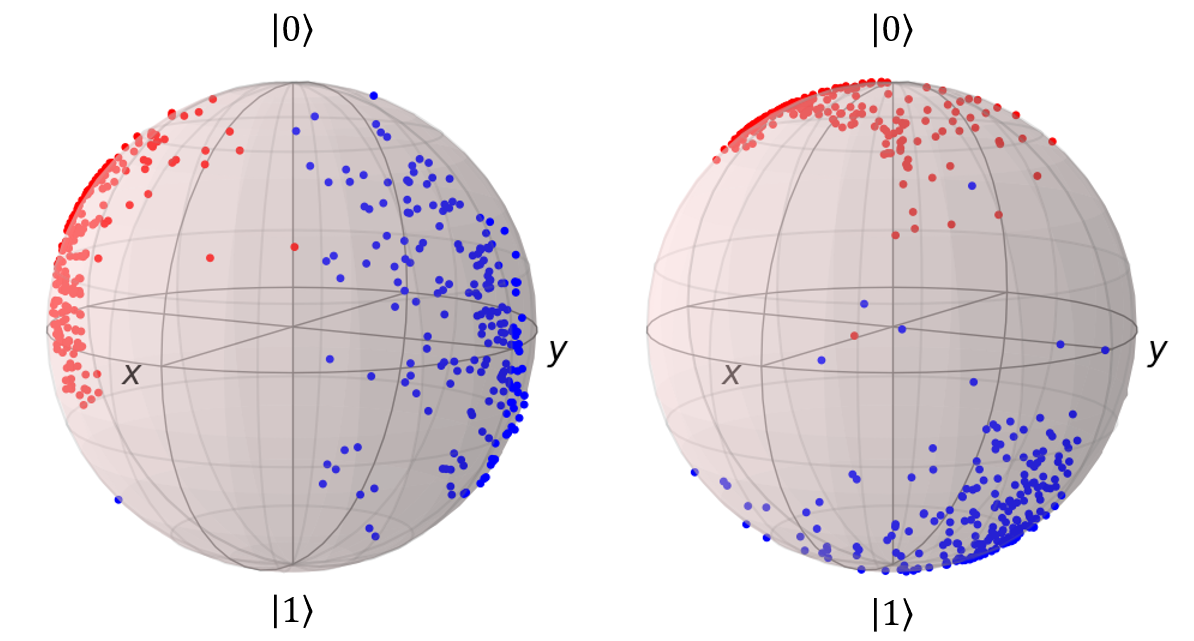}
    \caption{The left figure shows the results of feature extraction on Bloch spheres compressing 8 qubits into 1 qubit. The red and blue points represent handwritten digits 0 and 1, respectively. After applying $R_X$ and $R_Y$ rotations, the encoded results shown in the right figure are concentrated around $\lvert 0 \rangle$ and $\lvert 1 \rangle$, and then measured.}
    \label{fig:BlochSphere}
\end{figure}

Regarding the training results, since it is uncertain where the compressed state for 0 and 1 is located on the Bloch Sphere, directly measuring along the $z$-axis is impractical. Therefore, to confirm the performance of our result, we measure all three directions ($x$, $y$, and $z$) to determine the position of the current compressed state. Using an Support Vector Machine(SVM) \cite{svm}, we find a plane equation on the Bloch Sphere that separates 0 and 1. Subsequently, we apply rotations, as shown in Figure \ref{fig:BlochSphere}, such that this plane becomes perpendicular to the $z$-axis. This adjustment eliminates the need to measure all three directions to determine the current position of the compressed state in Hilbert space.

\subsection{Simulation}
\subsubsection{Unsupervised feature extraction}

Accuracy of Binary Classification $0$ vs $1$ across different architectures In Table \ref{tab:MainTest1}, "Half" refers to using only the encoder in QAE. The encoder architecture is the same as what we discussed in Section IV. This results show that the QCNN architecture achieved an average accuracy of 97.59 $\%$, significantly outperforming architectures A and B, which achieved 88.40 $\%$ and 88.78 $\%$, respectively. This highlights the substantial advantages of employing the QCNN architecture in QAE and demonstrates that QAE is an effective and promising approach to unsupervised feature extraction.

\begin{table}[h!]
    \centering
    \includegraphics[width=10.8cm]{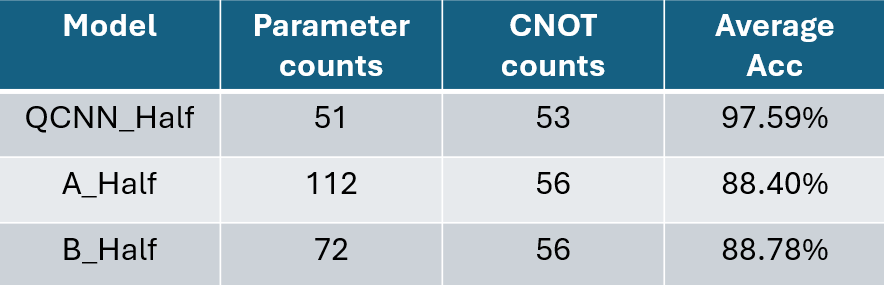}
    \caption{Comparison of QCNN, Architecture A, and Architecture B in Unsupervised Feature Extraction, Parameter Counts, and Number of CNOT Gates. To calculation Average Accuracy, we averaged the 8 training accuracy results.}
    \label{tab:MainTest1}
\end{table}

\subsubsection{Reconstruction Success Rate Across Different Architectures}
The results, shown in Table \ref{tab:MainTest2}, include both the encoder and decoder, which doubles the parameter count compared to the unsupervised feature extraction test. This evaluation aims to determine whether the QAE can achieve complete reconstruction.
The results indicate that QCNN performs slightly better than Architectures A and B, but the differences are not substantial. While QCNN demonstrated strong performance in unsupervised feature extraction, its reconstruction results are relatively underwhelming. This suggests room for improvement in developing more effective encoder structures or better reconstruction models in the future.

\begin{table}[h!]
    \centering
    \includegraphics[width=11.2cm]{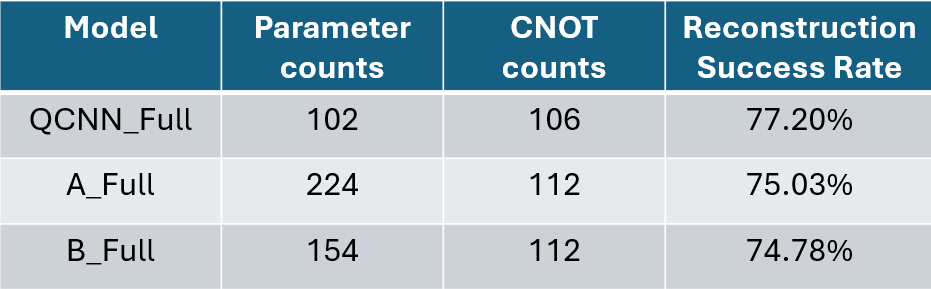}
    \caption{Comparison of Reconstruction Capabilities Between QCNN and Architectures A and B, Along with Their Parameter Counts and CNOT Gate Counts. Reconstruction success rate refers to the orthogonal relationship between the data after passing through the encoder and decoder and the original data, this formula can be written as $\bra{\mathcal{D}\big(\text{Tr}_t\big[\mathcal{E}(\ket{\phi(\vec{x})}, \theta_1)\otimes\ket{a}\big], \theta_2\big)}\ket{\phi(x)}$.}
    \label{tab:MainTest2}
\end{table}


\subsubsection{Other Category Tests}
Next, we tested additional categories to determine the most suitable approach for unsupervised feature extraction. These tests included comparisons of different encoding methods, loss functions, and numbers of qubits, all conducted using the QCNN architecture with a single compressed qubit.

As shown in Table \ref{tab:CompareTest1}, Angle Encoding outperformed Amplitude Encoding. This result contrasts significantly with the findings of T. Hur in \cite{aqcnn}.

\begin{table}[h!]
    \centering
    \includegraphics[width=8.4cm]{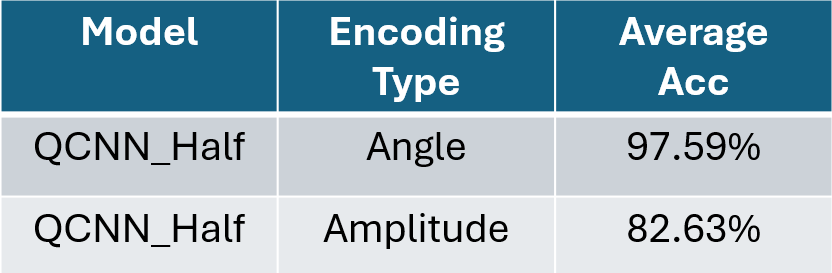}
    \caption{Accuracy of Different Embedding Types}
    \label{tab:CompareTest1}
\end{table}

The results, presented in Table \ref{tab:CompareTest2}, compare the performance of three pooling methods: no pooling, ZX Pooling, and Generalized Pooling. Additionally, we evaluated three convolutional structures, including the 3rd and 4th Convolutional Circuits proposed by T. Hur in \cite{aqcnn}, corresponding to the Type 2 and Type 3 architectures in the figure. The findings reveal that while there is little difference between ZX Pooling and Generalized Pooling, using any pooling method generally leads to better performance compared to no pooling. Among the convolutional structures, the first structure, Type 1, delivered the best results.

\begin{table}[h!]
    \centering
    \includegraphics[width=14cm]{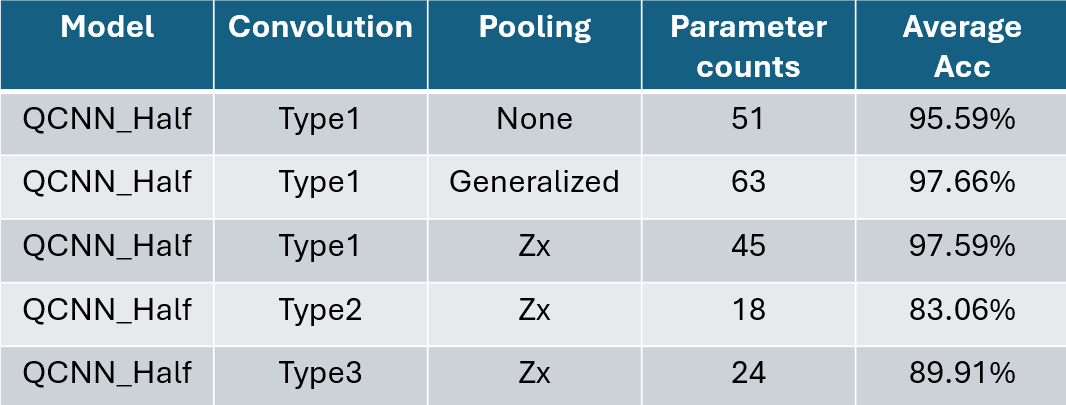}
    \caption{Accuracy of Different Pooling and Convolution Methods}
    \label{tab:CompareTest2}
\end{table}

We compared the performance of the MSE loss function with our primary choice, the BCE loss function. The results, presented in Table \ref{tab:CompareTest3}, are consistent with the findings of \cite{aqcnn}, demonstrating that MSE is more suitable than BCE as a loss function.
\begin{table}[h!]
    \centering
    \includegraphics[width=8.4cm]{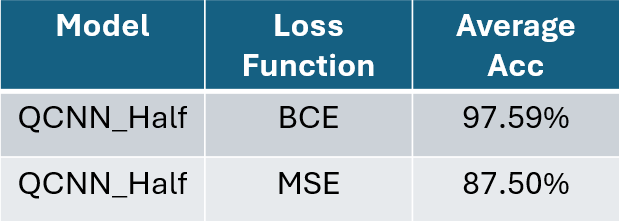}
    \caption{Comparision with MSE and BCE}
    \label{tab:CompareTest3}
\end{table}

We reduced 16-dimensional data to a 1-dimension and compared the performance of 8 qubits and 16 qubits. As shown in Table \ref{tab:CompareTest4}, the performance of 16 qubits was weaker than that of 8 qubits. This indicates that further methods will be required in the future to address this issue.
\begin{table}[h!]
    \centering
    \includegraphics[width=16.4cm]{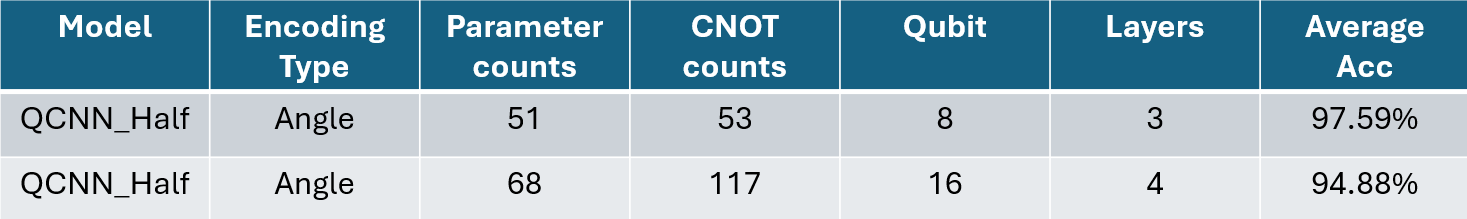}
    \caption{Comparison of 16 Qubits and 8 Qubits. 
    The circuit of 16 Qubits is "C-P-C-P-C-P-C-P" (see Figure \ref{fig:StructureOfQCNN}), using four layers of Convolution and Pooling.    }
    \label{tab:CompareTest4}
\end{table}

\subsubsection{Multi-Class Classification Results Across Different Layers}

In addition to binary classification of MNIST, we evaluated the performance on 3-class and 4-class classifications. For the 3-class classification, we used the digits 0, 1, and 8, while for the 4-class classification, we included the digits 0, 1, 2, and 3. The selection of 0, 1, and 8 is particularly noteworthy, as these digits differ in the number of "circles" (shapes) they contain.

\begin{figure}[h!]
    \includegraphics[width=16.4cm]{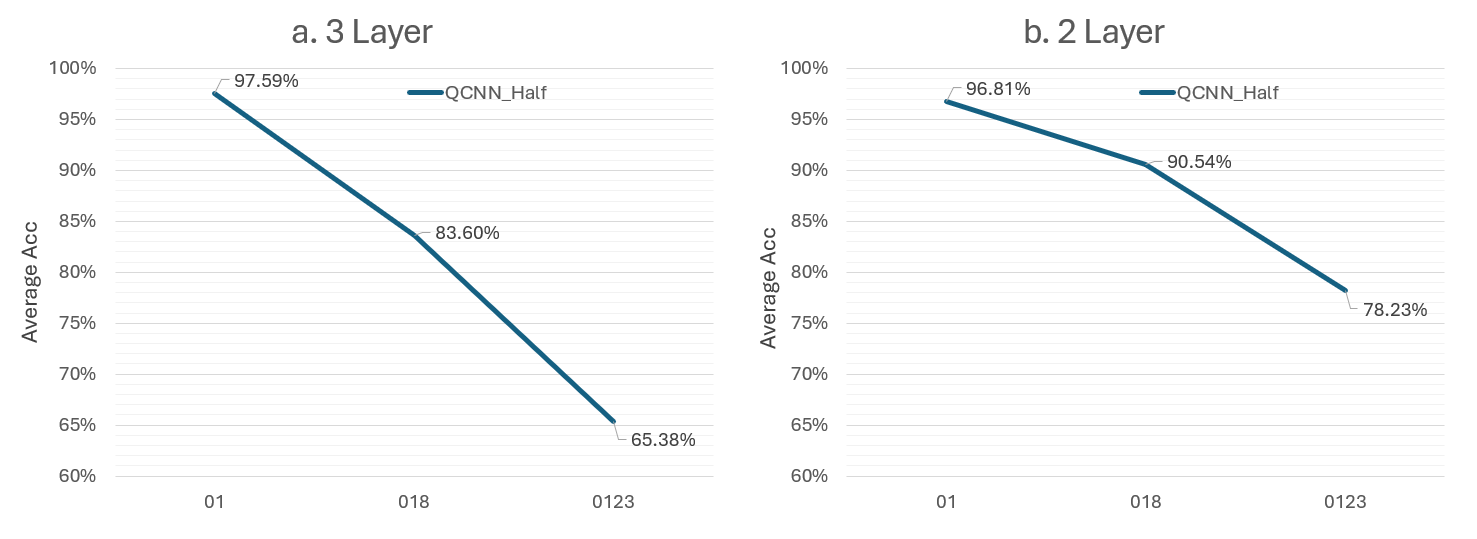}
    \caption{3 Layers (left) and 2 Layers (right): The x$-$axis represents the selected MNIST digits, and the y$-$axis represents accuracy. The left and right plots correspond to the 3 Layers(C-P-C-P-C-P) and 2 Layers(C-P-C-P) architectures (see Figure \ref{fig:StructureOfQCNN} for reference),  respectively, both tested using the QCNN Half and QCNN Full structures.}
    \label{fig:CompareTest5}
\end{figure}

As shown in Figure \ref{fig:CompareTest5}-a, the QAE's performance on 3-class and 4-class classifications was not very satisfactory. In comparison, M. Mordacci's results \cite{mor} using QCNN for supervised multi-class classification achieved an accuracy of 85\% for 4-class classification, significantly higher than our results. This disparity likely arises from the fundamental differences between unsupervised and supervised learning approaches, although the specific choice of digits might also contribute to the outcome.

Additionally, we simulated the performance of a reduced QCNN structure with only 2 layers, as illustrated in Figure \ref{fig:CompareTest5}-b. This structure includes two layers of convolution and pooling, producing an output of $\frac{8}{2^2} = 2$ compressed qubits. The reduced dimensionality compression ratio leads to a noticeable improvement in accuracy for both 3-class and 4-class classifications.

\subsubsection{Real simulation on Quantum Computer}

Finally, we evaluated the performance on a real quantum computer. Using the best results from the 16-qubit and 8-qubit simulations, we tested 100 MNIST data points (50 zeros and 50 ones) on the IonQ quantum computer using $1000$ shots. The results, summarized in Table \ref{tab:RealDevice1}, reveal that the performance on the IonQ quantum computer fell short of expectations.

\begin{figure}[h!]
    \includegraphics[width=16.4cm]{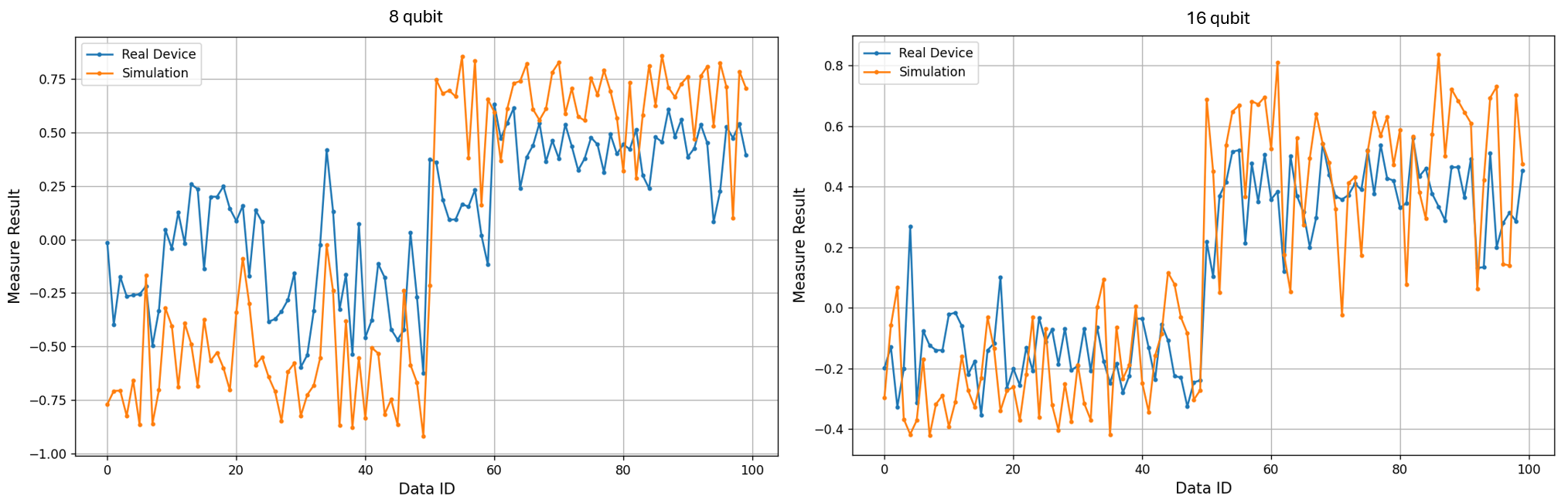}
    \caption{The results of quantum measurements with 8 qubits (left) / 16 qubit (right) and classical computer simulations are plotted, where Data IDs 0–49 correspond to handwritten digit 0, and Data IDs 50–99 correspond to handwritten digit 1.}
    \label{fig:RealDevice2}
\end{figure}

As illustrated in Figure \ref{fig:RealDevice2}, the results from the real device (IonQ quantum computer) showed a noticeable bias toward zero. This bias is likely attributable to the limitations of the current NISQ (Noisy Intermediate-Scale Quantum) era. Despite the use of higher-fidelity quantum wells, noise remains a significant factor, adversely affecting the accuracy of quantum computations.
\begin{table}[h!]
    \centering
    \includegraphics[width=16.4cm]{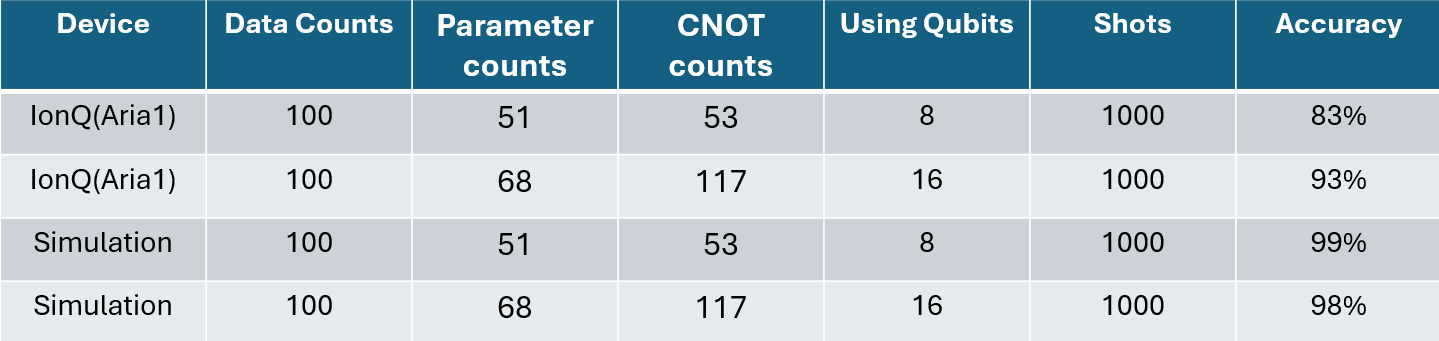}
    \caption{Comparison Between IonQ Quantum Computer Simulations and Classical Computer Simulations. } 
    \label{tab:RealDevice1}
\end{table}

\section{Conclusion}

We integrated the Quantum Autoencoder (QAE) with the QCNN architecture and compared its performance in unsupervised feature extraction and reconstruction against two alternative architectures. For unsupervised feature extraction on the MNIST dataset, the QCNN-based QAE achieved an accuracy of 97.59 $\%$, surpassing the other two architectures. This result highlights the effectiveness of QAE and the advantages of leveraging the QCNN framework. However, the reconstruction results in another test were suboptimal, prompting us to explore quantum circuit models or architectures better suited for reconstruction tasks.

Additionally, we examined various encoding methods and loss functions, concluding that angle encoding combined with binary cross-entropy loss is optimal for the Quantum Autoencoder. However, simulations with a 16-qubit setup yielded weaker results compared to the 8-qubit setup. Addressing this performance disparity is crucial for realizing the potential benefits of QCNN and QAE in larger qubit systems.

Finally, we implemented unsupervised feature extraction using the QAE with QCNN architecture on the IonQ quantum platform. Due to the limitations of current NISQ devices, the results on IonQ were less favorable than those obtained from simulations. Nevertheless, we are optimistic that advancements in quantum computing will enable our research on unsupervised feature extraction to achieve practical real-world applications.

\section{Data Availibility}
The dataset and code utilized in this study are publicly available and can be accessed at the following GitHub repository:
\url{https://github.com/Wattgo-Real/QAE-ClassificationMnist}

\section{Acknowledgments}
EJK (112-2124-M-002-003) and LYH (113-2112-M-033-006) acknowledge financial support from the National Science and Technology Council (NSTC) and the National Center for Theoretical Sciences in Taiwan.








\end{document}